\documentstyle[twocolumn,epsf,aps]{revtex}
\draft
\begin{document}

\title{Lattice-gas model for alkali-metal fullerides:\\
body-centered-cubic structure}
\author{Gy\"orgy Szab\'o}
\address
{Research Institute for Materials Science, H-1525 Budapest,
POB 49, Hungary}
\author{L\'aszl\'o Udvardi}
\address
{Quantum Theory Group, Institute of Physics, Technical
University of Budapest,\\
H-1111 Budapest, Budafoki \'ut 8, Hungary}
\address{\em \today}
\address{
\centering{
\medskip \em
\begin{minipage}{15.4cm}
{}~~~A Coulomb lattice-gas model with a host lattice screening mechanism is
adapted to describe the ordering phenomena in alkali-metal fullerides
of body-centered-cubic structure. It is assumed that the electric charge
of an alkali ion residing at a tetrahedral interstitial site is
completely screened by its first-neighbor C$_{60}$ molecules. The
electronic energy of the C$_{60}^{x-}$ ion is also taken into
consideration as a charged spherical shell. By means of these assumptions an
effective (short range) pair interaction is obtained between two alkali ions.
The resultant lattice-gas model is analyzed by using two- and six-sublattice
mean-field approximations. The thermodynamic properties are summarized
in phase diagrams for different shell radii.
\end{minipage}
}}
\maketitle
\narrowtext

\section{INTRODUCTION}
\label{sec:intro}

In a previous paper \cite{I} (henceforth referred to as I) we have
introduced a lattice-gas model to describe the ordering of alkaline
atoms intercalated into solid C$_{60}$ of face-centered-cubic (FCC)
structure. This model now is adapted to a body-centered-cubic (BCC) host
lattice because these structures are also observed in experiments
(see reviews in refs. \cite{fischer,weaver}).

Lattice-gas (Ising) models provide an adequate description of ordering
phenomena in intercalated alloys. Difficulties arise, however, when
determining the effective pair interaction between two intercalated
particles. In A$_x$C$_{60}$ alkali-fullerides the alkali atoms transfer
their $s$ electron to C$_{60}$ molecules. As a result the Coulomb energy
becomes dominant in the formation of different ordered
structures.\cite{fleming,rabe}  At the same time these transferred
electrons take part in the screening of Coulomb interaction between two
alkaline ions. In the knowledge of the screening mechanism one could
derive an effective pair interaction expected to be a short range one.
Unfortunately, we have no satisfactory approach for determining
the screening in these materials.

In the present model we assume a simple screening mechanism. First of all
the effect of dielectric media is considered viaa dielectric constant
$\varepsilon$. Furthermore, the
charge of each alkali ion is completely screened by the first neighbor
C$_{60}$ molecules. This means a uniform distribution of transferred $s$
electrons to the first neighbor C$_{60}$ molecules. This simplification
seems to be also an adequate approximation in a metallic system if the
Thomas-Fermi length is not greater than the shortest A-C$_{60}$ distance
($\approx 6.3$ \AA\ ).

In accordance with the above assumption the charge assembled on a given
C$_{60}$ molecule may not always the same: its value depends on the number of
alkali ions residing around the molecule. A series of UHF-MNDO
calculations has confirmed that the electronic energy of a solitary
C$_{60}^{x-}$ ion may be approximated as $E(x)=a+bx+cx^2$ where the
quadratic term mediates an interaction between those alkali ions
transferring charges to the same C$_{60}$ molecule(s). This contribution
may be considered as the electrostatic energy of a charged spherical
shell with a radius $R=4.45$ \AA\ because the term $a+bx$ does not
affect the thermodynamic behavior. The effect of dielectric media is
taken into account with the application of the Kirkwood-Onsager theory
\cite{ko} as in I. Rabe {\it et al.}
suggested such a spherical shell model choosing $R=3.5$ \AA\ (and $\varepsilon
=1$) which is equivalent to the shell radius formed by the carbon nucleus
of C$_{60}$ molecule.\cite{rabe} We use the concept of this spherical shell
model because of its descriptiveness, i.e. the electronic energy of
the C$_{60}^{x-}$ ion will be estimated as $E(x)=(xe)^2/2 \varepsilon R$.

Without repeating the formalism introduced in I, we now concentrate on
the mean-field analysis of the model using six- and two-sublattice
approximations. The results are summarized in phase diagrams calculated
for three different values of $R$. In the first case $R$ is estimated by
using the data of UHF-MNDO calculations. Accepting the suggestion of
Rabe {\it et al.}\cite{rabe}, in the second case, we have determined the
phase diagram for $R=3.5$ \AA . Finally we have chosen an intermediate
value for $R$ to display a phase diagram different from previous ones.

In the analysis of the above model we restrict ourselves to the rigid
BCC structure. Our main purpose is to analyze the general features of
the present model in the parameter region relevant for alkali-metal
fullerides. A more rigorous comparison of the states appearing in
FCC, BCC and body-centered-tetragonal (BCT) host lattices
\cite{zhu,stephens,poirier,pichler} goes beyond the scope of the present
work.

\section{THE MODEL}
\label{sec:model}

In BCC structures there are only tetrahedral interstitial sites for
the intercalated alkali ions. To describe the distribution of A$^+$ ions
we introduce a site variable $\eta_i$ which is 1 if the interstitial
site $i$ is occupied by an alkali ion and 0 for empty sites. The energy
for any configuration of alkali ions is given by the lattice-gas
Hamiltonian
\begin{equation}
H = \sum_i \epsilon \eta_i + {1 \over 2} \sum_{ij}
V_{ij} \eta_i \eta_j \;
\label{eq:Hamiltonian}
\end{equation}
characterized by the site energy $\epsilon$ and the effective pair
interaction $V_{ij}$. Both quantities are determined by assuming the
screening mechanism mentioned above.

\begin{figure}
\centerline{\epsfxsize=8cm
            \epsfbox{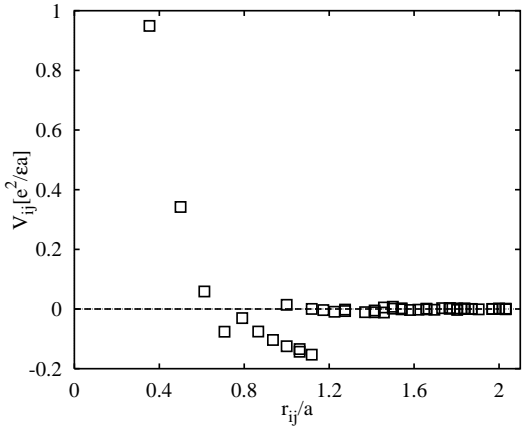}
            \vspace*{2mm}            }
\caption{Effective pair interaction vs. ion-ion distance for $R=4.45$
\AA .}
\label{fig:Vij}
\end{figure}

The calculation detailed in I results in an effective pair interaction
dependent on the ion-ion distance $r_{ij}$ as shown in
Fig.~\ref{fig:Vij} for $R=4.45$ \AA . Henceforth the energy is measured
in units of $e^2/(\varepsilon a)$ where $a$ is the lattice constant. Notice
that
$V_{ij}$ may have two values for certain values of $r_{ij}$. This
effective interaction is repulsive for short distances and becomes
attractive when increasing the ion-ion distance. In agreement with
expectations the strength of $V_{ij}$ decreases rapidly with $r_{ij}$.

The interaction mediated by  C$_{60}^{x-}$ ion(s) gives a repulsive
contribution to the total pair interaction until $r_{ij}$ exceeds a
threshold value. Obviously, the strength of this contribution increases
when decreasing $R$.

In the present model the site energy may easily be determined:
\begin{equation}
\varepsilon = -1.3752 {e^2 \over \varepsilon a}+{e^2 \over 8 \varepsilon R} \;
{}.
\label{eq:eps}
\end{equation}

In the knowledge of the model parameters we are able to determine the
energy of each ordered particle distribution. For this purpose the
tetrahedral interstitial sites are divided into 12 simple cubic
sublattices indicated by hexadecimal codes from 1 to C as illustrated in
Fig.~2.

\begin{figure}
\centerline{\epsfxsize=8cm
            \epsfbox{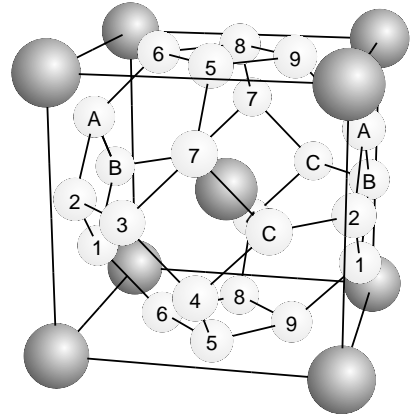}
            \vspace*{2mm}            }
\caption{Tetrahedral interstital sites of BCC lattice are divided into
12 sublattices.}
\label{fig:bcc}
\end{figure}

Within this sublattice division we can distinguish $2^{12}$ ordered states
in which the sites belonging to the same sublattice are uniformly
occupied or empty. Figure 3 shows the energies per C$_{60}$ of these
states as a function of alkali content ($x$). Here it is worth
mentioning that the charges of C$_{60}$ molecules are uniform within an
ordered state. Consequently, the energies plotted in Fig.~3 are the sums
of the corresponding Madelung energy and the estimated electronic energy
of C$_{60}$.

\begin{figure}
\centerline{\epsfxsize=8cm
            \epsfbox{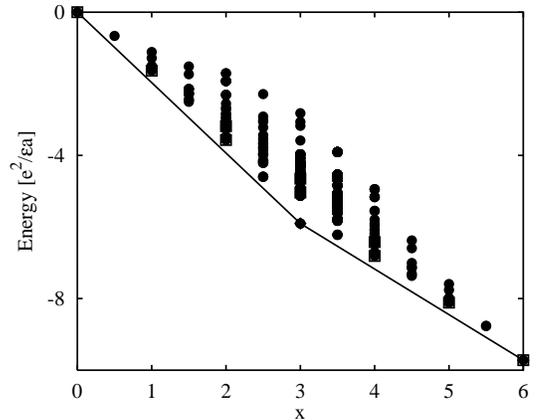}
            \vspace*{2mm}            }
\caption{Energy of 12-sublattice ordered states as a function of alkali-
metal content for $R=4.45$ \AA . Open diamonds and squares refer to
states described by the two- and six-sublattice formalisms. The solid
line connects the stable phases.}
\label{fig:gsen}
\end{figure}

Figure 3 demonstrates clearly that this model has only three stable
ordered structures if $R=4.45$ \AA . For the stable $x=0$ ($6$) state
all the tetrahedral sites are empty (occupied). In the half-filled
system ($x=3$) the stable state is equivalent to the well known A15
structure, i.e. the sublattices 2, 4, 6, 7, 9 and B are completely
occupied whereas the rest sites are empty. If the equivalent sublattices
are combined the formation of this structure may be well described by
using a much simpler two-sublattice formalism.

The states with the minimum energy for $x=1,2,4$ and $5$ exhibit additional
symmetry, more precisely, the occupations in sublattices $\alpha$ and
$\alpha+6$ are equivalent for $\alpha=1,...,6$. The combination of the
equivalent sublattices results in the six-sublattice formalism studied
previously.\cite{AgI} In this case the tetrahedral sites are divided
into six interpenetrating BCC sublattices therefore the translation
symmetry of the host lattice is conserved in the mentioned states. The
role of these states becomes important when studying the system for
lower $R$.

The decrease of $R$ gives an $x$-dependent contribution to the energies
(see Fig.~3) without modifying the energy differences for fixed $x$. The
minimum energy states for $x=1$ and $5$ become stable if $R/a<0.3411$.
The simultaneous appearance of the stable states for $x=1$ and $5$ is a
consequence of the particle-hole symmetry. In the $x=1$ stable state
only the sites of one of the six sublattices are occupied. If
$R/a<0.3342$ then the system exhibits stable states for $x=2$ and $4$.
In the $x=2$ stable state the simultaneous occupation of the first
neighbor sites is excluded.

The above ground state investigations support that the subsequent
mean-field analysis may be restricted to the two- and six-sublattice
approximations.

\section{MEAN-FIELD APPROXIMATION}
\label{sec:mf}

The six-sublattice mean-field approximation was previously introduced to
describe the ordering processes in superionic AgI where the I$^-$ ions
form a BCC cage lattice for the mobile Ag$^+$ ions.\cite{AgI} This
sublattice division is obvious because the number of tetrahedral
interstitial sites is just six times greater than the points in the BCC
host lattice. In this approach the equilibrium states are characterized
by the average sublattice occupations $\sigma_{\nu}$ ($\nu=1,...,6$).
The energy per C$_{60}$ molecule is expressed by these quantities as
\begin{equation}
{\cal H}=\sum_{\nu} \varepsilon \sigma_{\nu} + {1
\over 2} \sum_{\nu, \tau} J_{\nu \tau}\sigma_{\nu}
\sigma_{\tau}
\label{eq:Hmf}
\end{equation}
where $\varepsilon$ is defined by Eq.~(\ref{eq:eps}) and the mean-field
coupling constants $J_{\nu \tau}$ summarize the effective interactions
($V_{ij}$) between a site of sublattice $\nu$ and all of the sites
belonging to sublattice  $\tau$. Due to the symmetries of this
sublattice structure\cite{AgI} there are only three independent
mean-field coupling constants. The numerical calculations yield
\begin{eqnarray}
J_{11}&=&...=J_{66}= -1.16445 {e^2 \over \varepsilon a} \;, \nonumber \\
J_{12}&=&...=J_{65}= +0.08251 {e^2 \over \varepsilon a} \;,
\label{eq:Jnum} \\
J_{14}&=&...=J_{63}= -0.30468 {e^2 \over \varepsilon a} \;, \nonumber
\end{eqnarray}
if $R=4.45$ \AA .

In the second approach we distinguish only two sublattices ($\alpha$ and
$\beta$) as described in the previous section. Now the mean-field energy
per C$_{60}$ molecule is defined by the following expression:
\begin{equation}
{\cal H}=3\sum_{r} \varepsilon \sigma_{r} + {3
\over 2} \sum_{r, s} J_{r s}^{\prime}\sigma_{r}\sigma_{s}
\label{eq:Hmfp}
\end{equation}
where $r,s=\alpha , \beta$ and the values of $J_{r s}^{\prime}$ are
determined numerically for the same radius:
\begin{eqnarray}
J_{\alpha\alpha}^{\prime}&=&J_{\beta\beta}^{\prime}= -1.83360 {e^2 \over
\varepsilon a} \;, \nonumber \\
J_{\alpha\beta}^{\prime}&=&J_{\beta\alpha}^{\prime}= +0.69451 {e^2 \over
\varepsilon a} \;.
\label{eq:Jnump}
\end{eqnarray}
Within this formalism the twofold degenerate A15 structure is given as
$\sigma_{\alpha}=(1-\sigma_{\beta})=0$ or 1. Obviously, the energies of
the empty ($\sigma_{\alpha} = \sigma_{\beta}=0$) and completely occupied
($\sigma_{\alpha} = \sigma_{\beta}=1$) states are equivalent to those
suggested by the six-sublattice formalism.

Using the same method one can also evaluate the mean-field coupling
constants of the 12-sublattice approximation. The values of these parameters
are, of course, strongly related to the previous ones defined by
Eqs. (\ref{eq:Jnum}) and (\ref{eq:Jnump}). More precisely, $J_{\nu
\tau}$ and $J_{r s}^{\prime}$ are the linear combinations of the
mean-field coupling constants of the 12-sublattice approximation.

The equilibrium state characterized by sublattice occupations is
determined by minimizing the Gibbs potential for fixed temperature and
chemical potential. Following I, the results of this mean-field analysis
are summarized in phase diagrams.

In agreement with the particle-hole symmetry the phase diagrams are also
symmetric, as illustrated in Fig.~4. In this figure the dashed line
represents the critical temperature of the continuous sublattice
ordering from the high temperature ($\sigma_{\alpha}=\sigma_{\beta}$)
distribution towards the A15 structure. Below the critical temperature
$\sigma_{\alpha}$ differs from $\sigma_{\beta}$ and the alkali content
per C$_{60}$ molecule is given as $x= 3 (\sigma_{\alpha} +
\sigma_{\beta})$. In the subsequent phase diagrams this state is denoted
as A referring to the symmetry of A15 structure.

\begin{figure}
\centerline{\epsfxsize=8cm
            \epsfbox{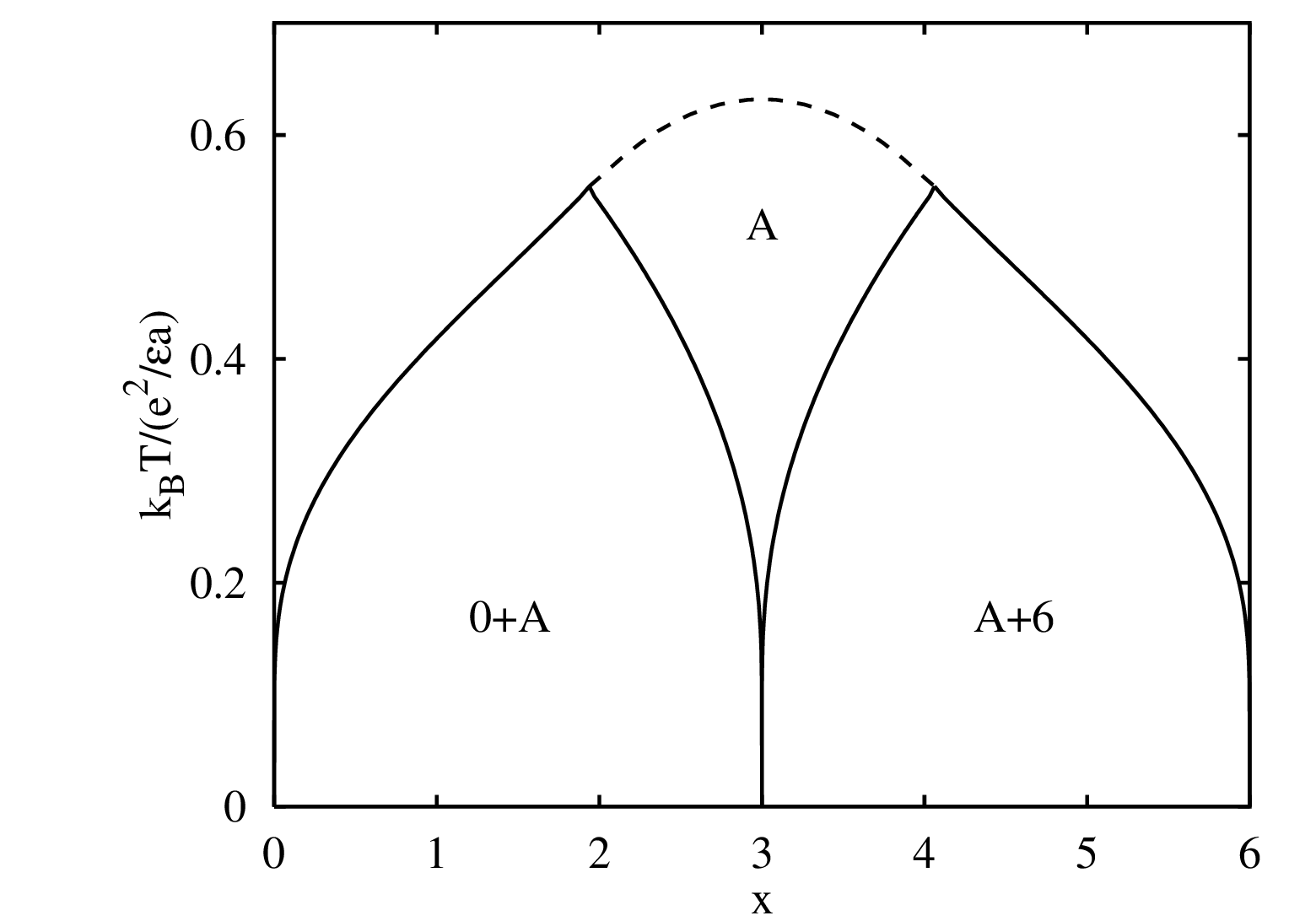}
            \vspace*{2mm}            }
\caption{Phase diagram suggested by mean-field approximation for
$R=4.45$ \AA\ . The dashed line represents a continuous two-sublattice
ordering process. The phases are indicated by the alkali content of
ordered states.}
\label{fig:pd385}
\end{figure}

The two-sublattice ordering process is analogous to the formation of
antiferromagnetic order in Ising models. The critical (N\'eel)
temperature may be expressed by the coupling constants as a function of
$x$, that is,
\begin{equation}
k_BT_N={x(6-x) \over 36}(J_{\alpha\beta}^{\prime}-
J_{\alpha\alpha}^{\prime}) \; .
\label{eq:TN}
\end{equation}
This transition temperature has a maximum at $x=3$ and vanishes when $x$
goes to 0 or 6. Accepting the numerical value $\varepsilon=9$ deduced
by Sanguinetti and Benedek \cite{sb} from the charge-induced vibrational
shift the above expression predicts $T_N \approx 1000$K for $x=3$. Instead
of this continuous transition, at low
temperatures the system segregates into two phases with different alkali
contents (indicated by solid lines in Fig.~4). The values refer to the
alkali contents in the ordered states at $T=0$.

We have determined the phase diagram for $R=3.5$ \AA\ ($R/a=0.303$)
suggested by Rabe {\it et al}.\cite{rabe} In this case only half
($3\leq x \leq 6$) of the symmetric phase diagram is represented in
Fig.~5.

\begin{figure}
\centerline{\epsfxsize=8cm
            \epsfbox{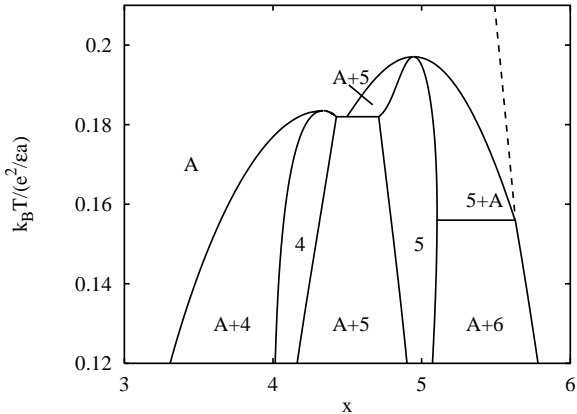}
            \vspace*{2mm}            }
\caption{Phase diagram in mean-field approximation for
$R=3.5$ \AA\ exhibits a eutectoid phase transition at $x\approx 4.5$\ .}
\label{fig:pd303}
\end{figure}

This figure shows stable ordered structures for $x=3$, 4, 5, and 6 at
low temperatures. The $x=3$ state transforms continuously to the random
distribution with a critical temperature given by Eq. (\ref{eq:TN}). A
part of this curve is denoted by the dashed line in Fig.~5. As demonstrated,
state A can exist for a wide range of $x$. The ordered structures
with nominal compositions $x=4$ and 5 become unstable when increasing
the temperature above $k_BT_4=0.1835 e^2/(\varepsilon a)$ and $k_BT_5=0.1971
e^2/
(\varepsilon a)$. Furthermore the model exhibits a eutectoid phase transition
at
$x \approx 4.5$. This phenomenon implies the stability of an ordered
structure with $x=4.5$ for smaller $R$.

Here it is worth mentioning that the cubic symmetry is broken in the
ordered structures if $x=1$, 2, 4 and 5. As a consequence, the diffusion
becomes anisotropic if the particles (holes) reside predominantly in two
(second neighbor) sublattices.\cite{sk} It seems reasonable to suggest
that a distortion of the host lattice occurs because of the breaking of
cubic symmetry related to the ordering process. A similar phenomenon is
investigated by O'Sullivan {\it et al.} with MD simulations when
studying the ordering process in the superionic phase of
AgI.\cite{MDAgI}

As mentioned above there exists a very narrow range of $R$ where the
ordered state $x=5$ is observable whereas the state $x=4$ is unstable.
In order to illustrate typical behavior within this region we have
evaluated the phase diagram for $R/a=0.337$.

In Fig.~6 the dashed line refers to the critical temperature of the
continuous sublattice ordering described above. In this phase diagram
there is a region denoted as A+A' where two A phases coexist with
different compositions. At lower temperatures (see region A+6) the
difference between $\sigma_{\alpha}$ and $\sigma_{\beta}$ disappears for
large alkali content ($x\approx 5.5)$. State 5 (with five of six
occupied sublattices) becomes stable if $k_BT<0.1615 e^2/(\varepsilon a)$.

\begin{figure}
\centerline{\epsfxsize=8cm
            \epsfbox{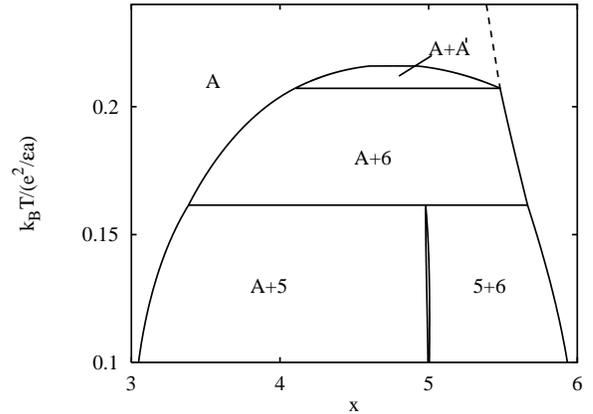}
            \vspace*{2mm}            }
\caption{Phase diagram for  $R/a=0.337$\ .}
\label{fig:pd337}
\end{figure}

The above series of phase diagrams visualizes the electronic energy effect
of the C$_{60}^{x-}$ ion on the thermodynamic behavior. In contrast
to the previous experiments and calculations the present model suggests
the appearance of state 5 if state 4 is stable. Very recently, however,
Lof {\it et al.} \cite{lof}
have found a local minimum near $x=5$ when measuring the temperature
and concentration dependent conductivity of potassium doped C$_{60}$
films. This observation may be interpreted as an experimental indication
for the existence of such a state. Here it is emphasized, that this state
is sixfold degenerated therefore a polydomain structure is expected to
appear in experiments.

\section{SUMMARY AND CONCLUSIONS}
\label{sec:conc}

Using two- and six-sublattice mean-field approximations we have studied
a lattice-gas model of intercalation alloys in which the Coulomb
interaction between the intercalated elements is screened out by the
charges distributed on the BCC host lattice. We have assumed that the
intercalated particle residing in a tetrahedral interstitial position
transfers $e/4$ charges to the (four) nearest neighbor host lattice
molecules. The energy of the charged host lattice molecule is treated as
the electrostatic energy of a charged spherical shell of radius $R$. The
family of alkali-fullerides is the best candidate to satisfy the above
conditions. This fact has motivated the choice of parameters when
determining the thermodynamic behavior suggested by this model. Although
our analysis is restricted to alkali-fullerides we can derive predictions
for earth alkalines if $2e$ is sunstituted for $e$ in all the formulae
including the energy unit in the figures.

The model takes the electrostatic interactions exactly into
consideration at zero temperature, i.e. it reproduces the Madelung
energies for the ordered structures. Owing to its simplicity the model has
only two parameters: the lattice constant is taken from experiments, the
value of $R$ is estimated by a series of UHF-MNDO calculations. In this
case the model suggests only three stable ordered structures. Besides the
empty ($x=0$) and fully occupied ($x=6$) states we have found a stable
A15 structure ($x=3$). The fully occupied ($x=6$) state is observed
experimentally for A=K, Rb and Cs. For smaller alkali elements (Li, Na),
however,  the interstitial voids are too large compared with ionic
sizes. Yildirim {\it et al.} suggest clusters of four to nine sodium
atoms inside the octahedral voids of the FCC lattice.\cite{yild} The
structure of Na$_6$C$_{60}$ exhibits an FCC host lattice with single
occupation of tetrahedral sites and fourfold filling of the octahedral
voids.\cite{ross} Obviously, the description of these former structures
goes beyond the validity of the present lattice-gas formalism.

According to the electrostatic force calculations the FCC structure is
preferable to the BCC one for $x \leq 3$.\cite{fleming,rabe} This theoretical
prediction agrees with experiments.\cite{fischer,weaver} Up till now only
the Ba$_3$C$_{60}$ compound has been found to exhibit the A15 structure.
\cite{kortan}

In the present model the robust A (A15) state can exist in a wide range
of temperature and alkali content as indicated in the phase diagrams. On
the analogy of antiferromagnetic ordering this particle arrangement
transforms continuously to the random distribution when the temperature
is increased.

The most surprizing prediction of the present calculation is the
existence of the A$_5$C$_{60}$ phase when the A$_4$C$_{60}$ compound
is stable. In these ordered structures the cubic symmetry is broken
therefore lattice distortions are expected to accompany the ordering
processes. For example, the A$_4$C$_{60}$ compounds have been observed
with BCT structure\cite{fleming} wich may be considered as a
distorted BCC lattice with four of six sublattices to be occupied. The
present model permits the existence of the stable A$_4$C$_{60}$ compounds
together with A$_5$C$_{60}$ if $R/a<0.334$. The $x=5$ composition,
however, has never been detected in experiments. The breaking of cubic
symmetry in these former states raises many questions related to the
role of lattice distortion. The investigation of the present model making
use of a BCT structure is in progress. The results of this analysis are
needed to enable a complete picture to be drawn.

\acknowledgements

We wish to acknowledge support from the Hungarian
National Research Fund (OTKA) under Grants Nos. 2960 and
F14378.


\begin{references}

\bibitem{I}L. Udvardi and G. Szab\'o, E-print: cond-mat/9512001

\bibitem{fischer}J. E. Fischer and P. A. Heiney, J.\ Phys.\
Chem.\ Sol.\ {\bf 54}, 1725 (1993).

\bibitem{weaver}J. H. Weaver and D. M. Poirier, Solid State
Physics {\bf 48}, 1 (1994).

\bibitem{fleming}R. M. Fleming, M. J. Rosseinsky, A. P.
Ramirez, D. W. Murphy, J. C. Tully, R. C. Haddon, T.
Siegrist, R. Tycko, S. H. Glarum, P. Marsh, G. Dabbagh, S.
M. Zahurak, A. V. Makhija, and C. Hampton,  Nature {\bf
352}, 701 (1991).

\bibitem{rabe}K. M. Rabe, J. C. Phillips, and J. M.
Vandenberg, Phys.\ Rev.\ B {\bf 47}, 13\,067 (1993).

\bibitem{ko}See, for example, C. J. Bottcher, {\it Theory of Electric
Polarization}, (Elsevier, Amsterdam, 1952).

\bibitem{zhu}Q. Zhu, O. Zhou, N. Coustel, G. M. M. Vaughan,
J. P. McCauley, Jr., W. J. Romanow, J. E. Fischer, A. B.
Smith, Science {\bf 254}, 545 (1991).

\bibitem{stephens}P. W. Stephens and L. Mihaly, Phys.\ Rev.\
B {\bf 45}, 543 (1992).

\bibitem{poirier}D. M. Poirier, T. R. Ohno, G. H. Kroll, P. J. Benning,
F. Stepniak, J. H. Weaver, L. P. F. Chibante, and R. E. Smalley, Phys.\
Rev.\ B {\bf 47}, 9870 (1993).

\bibitem{pichler}T. Pichler, R. Winkler, and H. Kuzmany, Phys.\ Rev.\ B
{\bf 49}, 15\,879 (1994).

\bibitem{AgI}G. Szab\'o, J.\ Phys. C {\bf 19}, 3775 (1986).

\bibitem{sb}S. Sanguinetti and G. Benedek. Phys.\ Rev.\ B {\bf 50},
15\,439 (1994).

\bibitem{sk}G. Szab\'o and J. Kert\'esz, J.\ Phys. C {\bf 19}, L273
(1986).

\bibitem{MDAgI}K. O'Sullivan, G. Chiarotti and P. A. Madden, Phys. Rev.
B {\bf 43}, 13\,536 (1991).

\bibitem{lof}R. W. Lof, H. T. Jonkman, and G. A. Sawatzky, Solid State
Commun.\ {\bf 93}, 633 (1995).

\bibitem{yild}T. Yildirim, O. Zhou, J. E. Fischer, N. Bykovetz, R. A.
Strongin, M. A. Cichy, A. B. Smith III, C. L. Lin, and R. Jelinek,
Nature {\bf 360}, 568 (1992).

\bibitem{ross}M. J. Rosseinsky, D. W. Murphy, R. M. Fleming,
R. Tycko, A. P. Ramirez, T. Siegrist, G. Dabbagh, and S. E.
Barrett, Nature {\bf 356}, 416 (1992).

\bibitem{kortan} A. R. Kortan, N. Kopilov, R. M. Fleming, O. Zhou, F. A.
Thiel, R. C. Haddon, and K. M. Rabe, Phys. Rev. B {\bf 47}, 13\ 070
(1993).

\end{references}
\end{document}